\documentclass[conference]{IEEEtran}
\IEEEoverridecommandlockouts
\usepackage{subcaption}

\usepackage{cite}
\usepackage{amsmath,amssymb,amsfonts}
\usepackage{algorithmic}
\usepackage{graphicx}
\usepackage{textcomp}
\usepackage{xcolor}
\usepackage{hyperref}
\usepackage{float}
\usepackage{multirow}
\usepackage{url}      

\usepackage{subcaption}

\def\BibTeX{{\rm B\kern-.05em{\sc i\kern-.025em b}\kern-.08em
    T\kern-.1667em\lower.7ex\hbox{E}\kern-.125emX}}
    
\begin{document}

\title{DiffEditor: Enhancing Speech Editing with Semantic Enrichment and Acoustic Consistency\\
\thanks{\IEEEauthorrefmark{1} Yong Qin is the corresponding author.}

}



\author{\IEEEauthorblockN{Yang Chen$^{1}$, Yuhang Jia$^{1}$, Shiwan Zhao$^{1}$, Ziyue Jiang$^{2}$, Haoran Li$^{1}$, Jiarong Kang$^{1}$ and Yong Qin$^{1,*}$}
\IEEEauthorblockA{$^1$\textit{College of Computer Science, Nankai University}, Tianjin, China, \\
$^2$\textit{Zhejiang University}, Zhejiang, China\\
Email: 2120230617@mail.nankai.edu.cn,
qinyong@nankai.edu.cn}
}


\maketitle

\begin{abstract}
As text-based speech editing becomes increasingly prevalent, the demand for unrestricted free-text editing continues to grow. However, existing speech editing techniques encounter significant challenges, particularly in maintaining intelligibility and acoustic consistency when dealing with out-of-domain (OOD) text. In this paper, we introduce \textit{DiffEditor}, a novel speech editing model designed to enhance performance in OOD text scenarios through semantic enrichment and acoustic consistency. To improve the intelligibility of the edited speech, we enrich the semantic information of phoneme embeddings by integrating word embeddings extracted from a pretrained language model. Furthermore, we emphasize that inter-frame smoothing properties are critical for modeling acoustic consistency, and thus we propose a first-order loss function to promote smoother transitions at editing boundaries and enhance the overall fluency of the edited speech. Experimental results demonstrate that our model achieves state-of-the-art performance in both in-domain and OOD text scenarios. 
The code and demo are available at \href{https://github.com/NKU-HLT/DiffEditor}{https://github.com/NKU-HLT/DiffEditor}.

\end{abstract}


\section{Introduction}
The advent of digital media and the rapid expansion of internet technologies have significantly transformed how we create and share content. Speech editing, a technique that modifies audio by altering its corresponding text, has become increasingly relevant in this context. Speech editing \cite{tan2021editspeech, campnet,voicecraft,instruct,le2024voicebox} enables users to edit the transcript rather than the audio itself, streamlining the process of generating and refining spoken content. This approach has become particularly valuable in diverse applications such as video production for social media, online course development, and film dubbing.

Speech editing models primarily focus on enhancing both the intelligibility of the synthesized speech and the acoustic consistency of the output. Many studies have been dedicated to improving these two metrics. For example, A\(^{\mathrm{3}}\)T \cite{a3t} proposes an alignment-aware approach that reconstructs masked acoustic signals during training by using text inputs and acoustic-text alignments, significantly improving the intelligibility through correct phoneme duration. Moreover, FluentSpeech \cite{flu} employs a context-aware diffusion model \cite{ddpm,ddim,diffusion} and a pretrained MFA model \cite{mfa} to further improve the alignment and adds a variation adaptor \cite{fs2} to model the pitch, thereby remarkably enhancing both the intelligibility and acoustic consistency. Based on this, FluentEditor \cite{liu2023fluenteditor} incorporates one loss based on GST \cite{gst} and another loss that only considers the two adjacent frames at the edited boundaries to enhance acoustic consistency.

However, existing methods predominantly follow in-domain text testing and training paradigms, focusing on continuous spectrograms derived from high-quality datasets \cite{vctk,ljspeech17}. These datasets feature standardized text corpora, which limit their applicability to real-world scenarios where users often provide more diverse and flexible inputs, thus frequently resulting in speech outputs of OOD text with unnatural prosody and incorrect pronunciation \cite{multi_feature}.

To address the OOD text problem in speech editing, we propose DiffEditor, a novel approach consisting of two key components. The first component enhances the model's intelligibility for OOD text by incorporating prior knowledge with rich text coverage, mitigating issues such as unclear pronunciation and errors \cite{multi_feature}. The second component optimizes both the editing boundaries and overall fluency by leveraging inter-frame smoothing properties, ensuring acoustic consistency throughout the entire utterance.

Building on this foundation, we enhance intelligibility by integrating word embeddings \cite{wordemb} derived from a pretrained BERT \cite{bert} model. These embeddings, trained on extensive text corpora, effectively capture the semantic and contextual nuances of OOD text, thereby improving the intelligibility of the edited speech. Additionally, we recognize that inter-frame smoothing properties play a vital role in achieving acoustic consistency. To this end, we introduce a first-order difference loss function during model training. The unique characteristics of the first-order difference enable our DiffEditor to focus not only on the transitions between the reserved and generated audio at the boundaries but also on the pattern of inter-frame changes within the edited speech, ultimately enhancing the overall acoustic consistency of the entire utterance.

\begin{figure*}[t]  
    \centering  
    \includegraphics[height=7cm, width=1.02\textwidth]{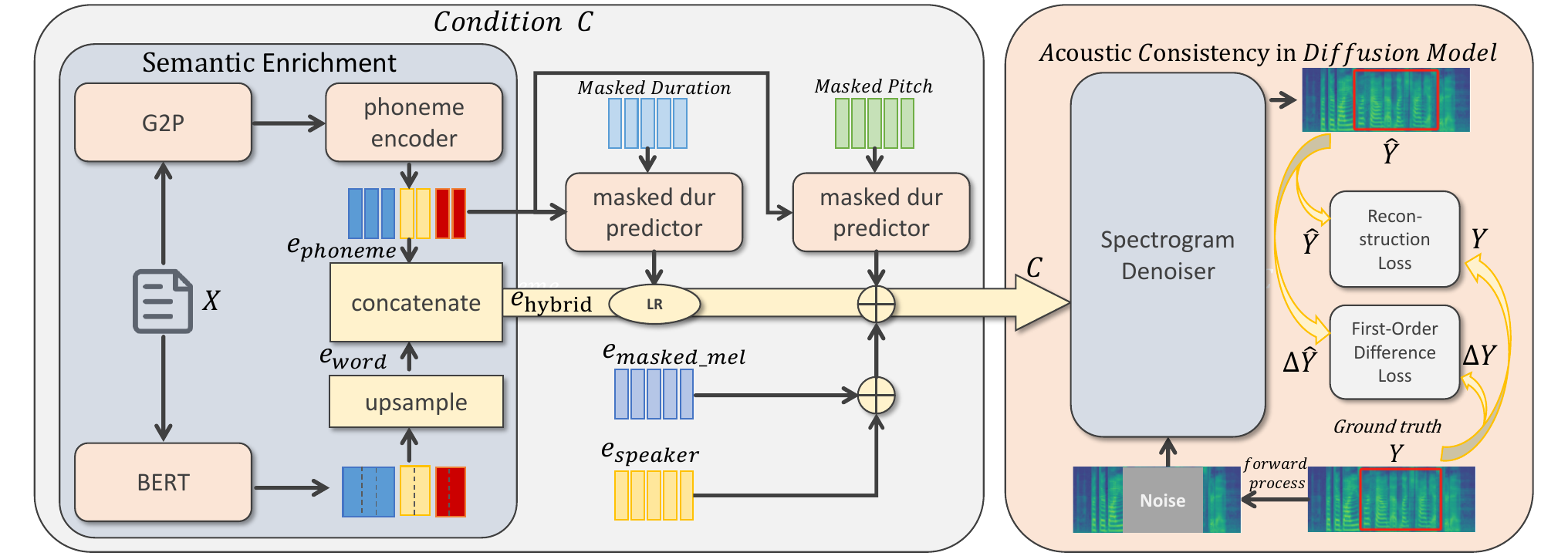}  
     \caption{Overview of the DiffEditor workflow. 1) The Condition C section illustrates the complete generation process for condition C, with the inner Semantic Enrichment section representing the Semantic Enrichment method. 2) The right section depicts the Diffusion Model, while the First-Order Difference Loss facilitates Acoustic Consistency.}  
    \label{fig:The overall workflow of DiffEditor.}  
\end{figure*}

Our subjective and objective experiments indicate that DiffEditor achieves state-of-the-art performance in both in domain and OOD text scenarios.

\section{METHODOLOGY}
We will first provide an overview of the DiffEditor workflow, followed by a detailed discussion of the specific implementations of both semantic enrichment and acoustic consistency.

\subsection{Overall Workflow}
As shown in \ref{fig:The overall workflow of DiffEditor.}, DiffEditor utilizes the Conditional Diffusion Model, which includes the condition generation module and the Spectrogram Denoiser module.

Assuming the edited input text is represented as \( X = (X_1, X_2, X_3, \ldots, X_{|x|}) \), it is processed through the G2P \cite{g2p} module to obtain the phoneme sequence denoted as \( P = (P_1, P_2, \ldots, P_{|p|}) \). After passing through the phoneme encoder, phoneme embeddings \( e_{phoneme} \) are generated. Meanwhile, the text \( X \) is processed through the BERT model to obtain word embeddings \( e_{word} \). These word embeddings are subsequently upsampled to align with the phoneme embeddings. Then, \( e_{word} \) is concatenated with the \( e_{phoneme} \) to produce \( e_{hybrid} \).  

The hybrid embedding \( e_{hybrid} \) are upsampled to frame-level features by utilizing a length regulator, which leverages the temporal information produced by the masked duration predictor. These frame-level features, along with additional components such as pitch \( e_{pitch} \), speaker embedding \( e_{speaker} \), and the masked mel-spectrogram embeddings \( e_{masked\_mel} \), are combined with the hybrid embedding to form the condition \( C \). The generation of other embedding follows the configurations established in \cite{flu}.  

The ground truth spectrogram is represented as \( Y = (Y_1, Y_2, Y_3, \ldots, Y_{|Y|}) \), while its masked counterpart is denoted as \( Y_{mask} = \text{mask}(Y) \). The masking function follows the specifications outlined in \cite{flu}.  

The Spectrogram Denoiser processes the condition features \( C \), which are generated from both \( Y_{mask} \) and the input \( X \) and the forwards-processed spectrogram \( Y \) as inputs to produce \( \hat{Y} \), where \( \hat{Y} \) maintains the unmasked part of \( Y_{mask} \) while reconstructing the masked portions of the spectrogram.


\subsection{Semantic Enrichment}

Given that word embeddings derived from BERT effectively capture the semantic and contextual nuances of OOD text, thereby enhancing the intelligibility of edited speech, we propose concatenating the BERT-generated word embeddings with phoneme embeddings to enrich the semantic information. This approach is illustrated in the Senmantic Enrichment section of \ref{fig:The overall workflow of DiffEditor.}.

The BERT model is a transformer-based model designed to understand the context of words in a sentence by looking at the words that come before and after them, which is a significant advancement over previous models that read text in a unidirectional manner. This training approach is very similar to the masked prediction method used in speech editing. Therefore, this model addresses the semantic information deficiency of phoneme embedding and enhances the understanding of contextual semantic information.

Following the processing of text through BERT, we upsample the word embeddings to align with the structure depicted by the dotted line in the Semantic Enrichment section of Figure \ref{fig:The overall workflow of DiffEditor.}. This ensures that the upsampled word embeddings can be seamlessly concatenated with the phoneme embeddings. Moreover, the timing of concatenation between \(e_{word}\)
  and \(e_{phoneme}\) 
  is critical; Random concatenation is insufficient. We outline three distinct concatenation strategies below, along with an analysis of their advantages and disadvantages.

First, we concatenate the phoneme embeddings and word embeddings to generate hybrid embeddings and use the hybrid embeddings as the input of the masked dur predictor and the masked pitch predictor. This approch will increase the predictors' parameters due to the expansion of embedding dimension and lead to the overfitting problem during training.

Second, we directly concatenate the word embeddings with the condition c. In this approach, all embeddings except the word embedding share the same dimension size of 192. However, global embeddings such as speaker embedding needs to function across the entire text representation, which is a concatenation of phoneme embeddings and word embeddings. Consequently, this type of model has lower performance.

Last but not least, we choose to concatenate \(e_{phoneme}\) and \(e_{word}\) immediately after the \(e_{phoneme}\) has passed through the masked duration predictor and masked pitch predictor. We then add the \(e_{masked_mel}\) and \(e_{speaker}\) to the \(e_{hybrid}\), as illustrated in the Condition C section of Figure \ref{fig:The overall workflow of DiffEditor.}. This strategy achieves the best performance in terms of semantic enrichment because it avoids the mistakes of the former two strategies.

\subsection{Acoustic Consistency}

We realize that the inter-frame smooth
properties play an important role in achieving acoustic consistency.
To this end, we introduce the first-order difference loss
during model training. 

The first-order difference \(\Delta Y\) for the ground-truth acoustic features \(Y\) is computed as:
\[
\Delta Y = Y^{(i)} - Y^{(i-1)}, \quad \text{for } i \geq 2
\]
where  denotes the index of the frame.
This calculation captures the rate of change between adjacent frames, particularly at the editing boundaries. To enforce temporal consistency and minimize abrupt changes in the predicted \( \hat{Y} \), we introduce the first-order difference loss \(\mathcal{L}_{FD}\). This loss compares the first-order differences of the predicted \( \hat{Y} \) with those of the ground-truth speech \( Y \) and is defined as:
\[
\mathcal{L}_{FD} = \text{MAE}(\Delta Y, \Delta \hat{Y})
\]
Here, \(\text{MAE}\) denotes the Mean Absolute Error. This loss is crucial for editing the boundaries of synthesized speech during speech generation. By enforcing \(\mathcal{L}_{FD}\), the model ensures that transitions between edited and unedited segments are natural and fluent, while preserving the overall fluency and naturalness of the generated audio.

\section{EXPERIMENTS AND RESULTS}
\subsection{Dataset}

We trained DiffEditor using the VCTK dataset \cite{vctk} and evaluated its performance employing both the VCTK and LJspeech test sets \cite{ljspeech17}. Both the VCTK and LJspeech are high-quality English Speech datasets. The VCTK Corpus consists of speech data spoken by 109 English speakers with various accents, and each speaker reads about 400 sentences from the same newspaper. As a result, we randomly chose 10 speakers as the test set for in domain text evaluation, and the remaining 99 speakers made up the training set.
According to the summary provided by the large language model, the VCTK dataset encompassed eight topics: Daily Life, Sports, Interpersonal Relationships, Work, Politics, Education, Art, and Natural Phenomena, whereas the first 1000 utterances of the LJspeech dataset mainly describe the Newgate Prison in London and the related situations. Thus, for the OOD evaluation, we utilized the first 1000 utterances from the LJspeech dataset, which belongs to a completely different text domain.

\subsection{Experimental Setup}
The configuration of the phoneme encoder, variance adaptor \cite{fs2} (which includes the masked duration and pitch predictors), and the spectrogram denoiser are consistent with those described in \cite{flu}. The pretrained word encoder in our study is the BERT model, specifically the bert-base-multilingual-uncased\footnote{\hypertarget{link1}\url{https://huggingface.co/google-bert/bert-base-multilingual-cased}} version released by Google on Hugging Face. In our methodology, we assign a weight of 4 to the loss function associated with the first-order difference. The batch size is configured to 90, and DiffEditor is trained for 200,000 steps using a single NVIDIA 3090 GPU.

\subsection{Evaluation Metrics}
For subjective evaluation, we utilize Mean Opinion Score (MOS) \cite{mos} to assess speech quality, FMOS \cite{liu2023fluenteditor} to evaluate fluency, and IMOS to measure intelligibility. These metrics provide a detailed understanding of the model's improvements in these aspects. Additionally, we conduct a corresponding ablation study using CMOS (comparative mean opinion score) \cite{mos}, CFMOS, and CIMOS. For objective evaluation, we employ Mel-Cepstral Distortion (MCD) \cite{mcd}, Short-Time Objective Intelligibility (STOI) \cite{stoi}, and Perceptual Evaluation of Speech Quality (PESQ) \cite{pesq}.

\subsection{Comparative Study}
We have conducted extensive comparative experiments, which include five  models. they are: 1) \textbf{CampNet} \cite{campnet} presents a context-aware mask prediction network to mimic the process of text-based speech editing. 2) \textbf{EditSpeech} \cite{tan2021editspeech} trains two traditional autoregressive text-to-speech (TTS) models, one of them is trained in a left-to-right manner, and the other is trained in a right-to-left way.  3) \textbf{A\(^{\mathrm{3}}\)T} \cite{a3t} proposes an alignment-aware approach that reconstructs masked acoustic signals during training by leveraging text inputs and acoustic-text alignments. 4) \textbf{FluentSpeech} \cite{flu} advances A\(^{\mathrm{3}}\)T \cite{a3t} further with a context-aware diffusion model that iteratively refines the modified mel-spectrogram, guided by contextual features. 5) \textbf{DiffEditor} integrates multi-level feature reinforcement and first-order difference loss.

\begin{table*}[t]\tiny
\caption{ Objective evaluation results of comparative study for out-of-domain and in domain text.}
\centering 
\resizebox{\textwidth}{!}{
\begin{tabular}{l|ccc|ccc}
\hline

\multirow{2}{*}{\textbf{Method}} & \multicolumn{3}{c|}{\textbf{Out-of-Domain Text}} & \multicolumn{3}{c}{\textbf{In Domain Text}} \\
                       & \textbf{MCD} (↓) & \textbf{STOI} (↑) & \textbf{PESQ} (↑)  &   \textbf{MCD} (↓) & \textbf{STOI} (↑) & \textbf{PESQ}(↑)    \\

\hline

CampNet (ICASSP'22) & 9.304  & 0.253 & 1.153  & 9.588 & 0.393 & 1.258   \\

EditSpeech (ASRU'21)&9.411 & 0.342	& 1.107  & 8.984 & 0.559 & 1.239  \\

A\(^{\mathrm{3}}\)T (ICML'22) &9.321  &	 0.344 &	1.110  & 8.189 & 0.643 & 1.326   \\

FluentSpeech (ACL'23)& 7.815	& 0.608	& 1.317  & 7.015  & 0.749 & 1.641   \\

\hline

\textbf{DiffEditor} & \textbf{7.440} &	\textbf{0.627} &	\textbf{1.376}	& \textbf{6.777}  & \textbf{0.755} & \textbf{1.701}  \\
\hline



\end{tabular}
\label{table:1}
}
\end{table*}

\subsection{Main Results}

    \textbf{Objective results:} In order to objectively assess the performance of the out-of-domain (OOD) text and in domain text in the comparative study, we randomly mask a continuous span of the audio segment, which corresponds to a continuous phoneme sequence based on the alignment file generated by MFA \cite{mfa}. Subsequently, we utilize the five models to generate the masked part of the audio and calculate MCD, PESQ, and STOI to fill the table  \ref{table:1}. According to the evaluation, our DiffEditor achieves state-of-the-art performance in both in domain and OOD text scenarios for all three objective metrics. Simultaneously, we can observe that the performance of the OOD text significantly decreases in various models compared to the in domain text, proving that existing methods predominantly follow in domain text testing and training paradigms, thus having limited capability for OOD text scenario.

    \textbf{Subjective results:} We selected 45 utterances from the VCTK test set for OOD text subjective evaluation and used ChatGPT to edit these texts with OOD content automatically. We carefully set the ChatGPT prompts to ensure that the edited texts are randomly generated OOD texts. At the same time, we invited 20 English speakers to evaluate these audios in multiple dimensions, including MOS \cite{mos}, FMOS \cite{liu2023fluenteditor}, and IMOS. According to the final results shown in \ref{table:2}, DiffEditor achieves the best results in speech quality, fluency, and intelligibility, indicating that DiffEditor  improves the intelligibility of the edited speech and the
    overall acoustic consistency of the entire utterance.

\begin{table}[t]
\caption{Subjective evaluation results for out-of-domain text of FluentSpeech and DiffEditor.}
\centering
\renewcommand{\arraystretch}{1.1}
\resizebox{0.5\textwidth}{!}{
\begin{tabular}{l|cccc}
\hline
\textbf{Method}  & \textbf{MOS} & \textbf{FMOS}  & \textbf{IMOS} \\
\hline

FluentSpeech   &  4.11 ± 0.03 & 4.12 ± 0.03 &  4.20 ± 0.03 \\

\textbf{DiffEditor}  & \textbf{4.43 ± 0.06} & \textbf{4.43 ± 0.03} &  \textbf{4.47 ± 0.03}\\
\hline
\end{tabular}
\label{table:2}
}
\end{table}

\begin{table}[t]
\caption{Objective and subjective results of ablation study.}
\centering
\renewcommand{\arraystretch}{1.1}
\resizebox{0.5\textwidth}{!}{
\begin{tabular}{l|ccc|c}
\hline
\textbf{Method} &  \textbf{C-MOS}  & \textbf{C-FMOS}  & \textbf{C-IMOS} &\textbf{MCD} (↓)  \\
\hline
\textbf{DiffEditor} &\textbf{ 0.00} & \textbf{0.00} & \textbf{0.00} & \textbf{6.777}   \\

w/o \(\mathcal{L}_{FD} \) & -0.24 & -0.20 & -0.15 & 6.858   \\

w/o \(e_{word}\) & -0.27 & -0.16 &-0.17 &  6.904  \\
 
\hline   
 
\end{tabular}
\label{table:3}

}
\end{table}

\subsection{Ablation Study}

To further explain the influence of each component, we have conducted ablation experiments on the two methods employed by DiffEditor, namely \textbf{\textit ``w/o \(\mathcal{L}_{FD}\)''} and \textbf{\textit ``w/o \(e_{word}\)''}. The \textbf{\textit ``w/o \(\mathcal{L}_{FD}\)''} means that the DiffEditor subtracts the first-order difference loss function. Table \ref{table:3} indicates that the subtraction of the first-order difference loss function results in a decline in fluency and intelligibility, with the decrease in fluency being more significant than that when subtracting the word embeddings. This implies that the first-order difference enhances the overall acoustic consistency of the entire utterance. On the other hand, the subtraction of the word embeddings leads to a major decrease in intelligibility, and the decrease in intelligibility is more significant than that when subtracting the first-order difference loss function. This indicates that the word embedding captures the semantic and contextual nuances of the OOD text, thereby improving the intelligibility of the edited speech.

\subsection{Editing Boundaries Analysis}
\begin{figure}[]
   
     \caption{Visualizations of the generated mel-spectrogram boundary by
 DiffEditor and FluentSpeech}
    \centering
     \label{fig:semantic}
    \includegraphics[width=\columnwidth,height=0.1\textheight]{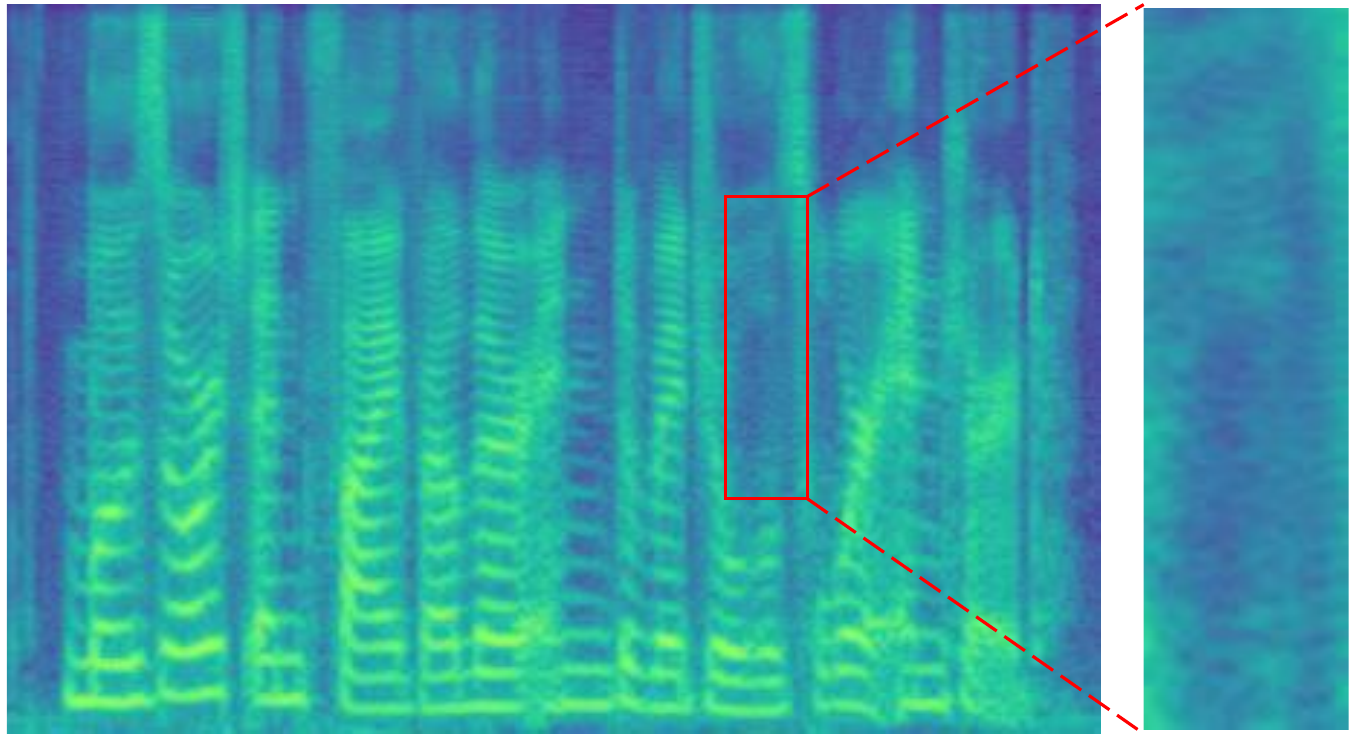} 
    \subcaption[a]{FluentSpeech}
    \label{fig:fluent_v}

    \includegraphics[width=\columnwidth,height=0.1\textheight]{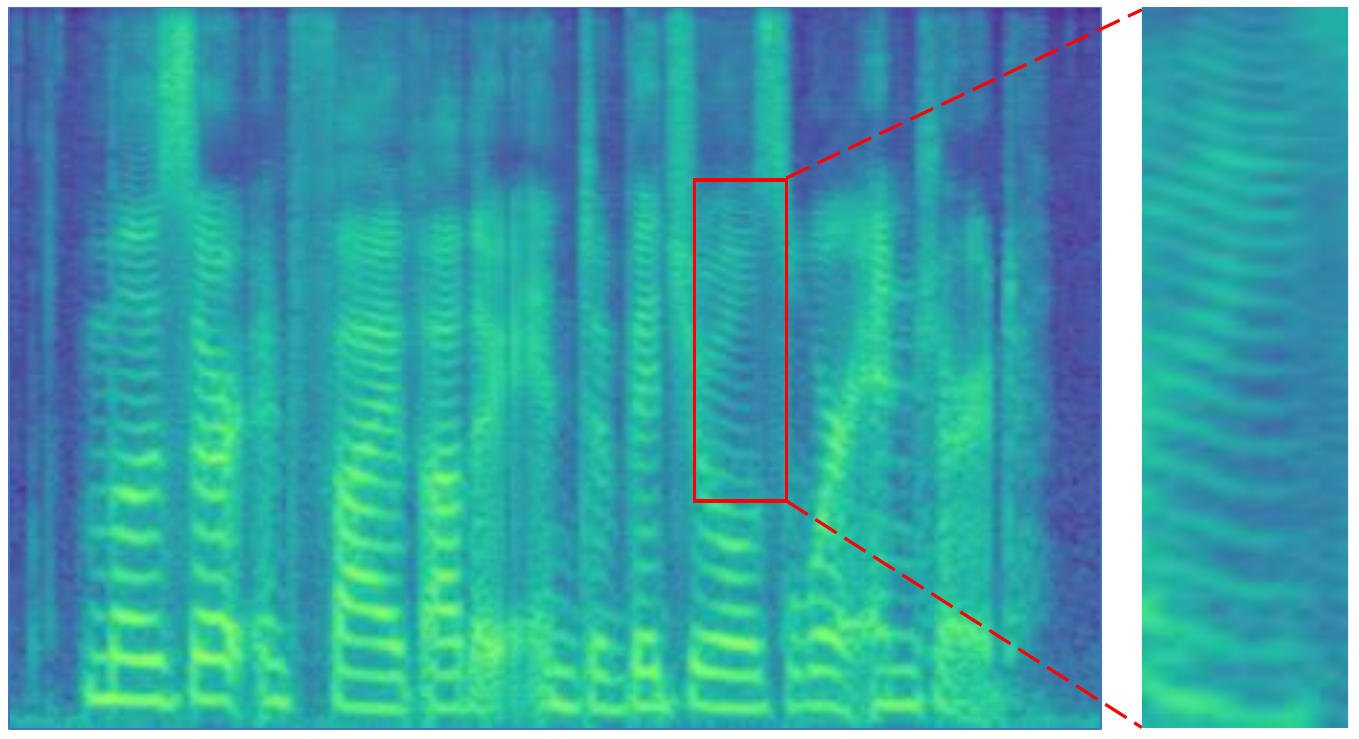}
    \subcaption[b]{DiffEditor}
    \label{fig:diff_v}

     \vspace{-15pt}

\end{figure}

As depicted in Figures \ref{fig:semantic}, we visualize the mel-spectrograms generated by the FluentSpeech baseline and our DiffEditor. The text for both spectrograms is edited from ``six other people were injure'' to ``mozart and five other individuals were injure''. The red box regions in the figures represent the precise editing boundary of the sentence, which is the generated ``individual''. As can be observed from the Figures \ref{fig:semantic}, when DiffEditor generates the boundaries, the spectrogram is more abundant and contains more detailed information, which demonstrates that the model learns the inter-frame transition pattern at the boundaries.

\section{CONCLUSION}

In this paper, we present DiffEditor, a novel speech editing framework that achieves state-of-the-art performance across both in domain and OOD text scenarios. By integrating word embeddings from a pretrained BERT model, we effectively incorporate prior knowledge with rich text coverage, significantly enhancing the model's intelligibility. Furthermore, by employing a first-order difference loss function, we ensure acoustic consistency at the editing boundaries and throughout the entire utterance.

\bibliographystyle{ieeetr}
\bibliography{ref}

\begin{thebibliography}{10}

\bibitem{tan2021editspeech}
D.~Tan, L.~Deng, Y.~T. Yeung, X.~Jiang, X.~Chen, and T.~Lee, ``Editspeech: A
  text based speech editing system using partial inference and bidirectional
  fusion,'' in {\em 2021 IEEE Automatic Speech Recognition and Understanding
  Workshop (ASRU)}, pp.~626--633, IEEE, 2021.

\bibitem{campnet}
T.~Wang, J.~Yi, R.~Fu, J.~Tao, and Z.~Wen, ``Campnet: Context-aware mask
  prediction for end-to-end text-based speech editing,'' {\em IEEE/ACM
  Transactions on Audio, Speech, and Language Processing}, vol.~30,
  pp.~2241--2254, 2022.

\bibitem{voicecraft}
P.~Peng, P.-Y. Huang, A.~Mohamed, and D.~Harwath, ``Voicecraft: Zero-shot
  speech editing and text-to-speech in the wild,'' {\em arXiv}, 2024.

\bibitem{instruct}
R.~Huang, R.~Hu, Y.~Wang, Z.~Wang, X.~Cheng, Z.~Jiang, Z.~Ye, D.~Yang, L.~Liu,
  P.~Gao, {\em et~al.}, ``Instructspeech: Following speech editing instructions
  via large language models,'' in {\em Forty-first International Conference on
  Machine Learning}.

\bibitem{le2024voicebox}
M.~Le, A.~Vyas, B.~Shi, B.~Karrer, L.~Sari, R.~Moritz, M.~Williamson,
  V.~Manohar, Y.~Adi, J.~Mahadeokar, {\em et~al.}, ``Voicebox: Text-guided
  multilingual universal speech generation at scale,'' {\em Advances in neural
  information processing systems}, vol.~36, 2024.

\bibitem{a3t}
H.~Bai, R.~Zheng, J.~Chen, M.~Ma, X.~Li, and L.~Huang, ``{A}$^3${T}:
  Alignment-aware acoustic and text pretraining for speech synthesis and
  editing,'' in {\em Proceedings of the 39th International Conference on
  Machine Learning}, vol.~162 of {\em Proceedings of Machine Learning
  Research}, pp.~1399--1411, PMLR, 17--23 Jul 2022.

\bibitem{flu}
Z.~Jiang, Q.~Yang, J.~Zuo, Z.~Ye, R.~Huang, Y.~Ren, and Z.~Zhao,
  ``Fluentspeech: Stutter-oriented automatic speech editing with context-aware
  diffusion models,'' in {\em Findings of the Association for Computational
  Linguistics: ACL 2023}, pp.~11655--11671, 2023.

\bibitem{ddpm}
J.~Ho, A.~Jain, and P.~Abbeel, ``Denoising diffusion probabilistic models,''
  {\em Advances in neural information processing systems}, vol.~33,
  pp.~6840--6851, 2020.

\bibitem{ddim}
A.~Q. Nichol and P.~Dhariwal, ``Improved denoising diffusion probabilistic
  models,'' in {\em International conference on machine learning},
  pp.~8162--8171, PMLR, 2021.

\bibitem{diffusion}
R.~Rombach, A.~Blattmann, D.~Lorenz, P.~Esser, and B.~Ommer, ``High-resolution
  image synthesis with latent diffusion models,'' in {\em Proceedings of the
  IEEE/CVF conference on computer vision and pattern recognition},
  pp.~10684--10695, 2022.

\bibitem{mfa}
M.~McAuliffe, M.~Socolof, S.~Mihuc, M.~Wagner, and M.~Sonderegger, ``Montreal
  forced aligner: Trainable text-speech alignment using kaldi,'' in {\em
  Interspeech 2017}, pp.~498--502, 2017.

\bibitem{fs2}
Y.~Ren, C.~Hu, X.~Tan, T.~Qin, S.~Zhao, Z.~Zhao, and T.-Y. Liu, ``Fastspeech 2:
  Fast and high-quality end-to-end text to speech,'' {\em Learning,Learning},
  Jun 2020.

\bibitem{liu2023fluenteditor}
R.~Liu, J.~Xi, Z.~Jiang, and H.~Li, ``Fluenteditor: Text-based speech editing
  by considering acoustic and prosody consistency,'' {\em Proc.
  InterSpeech2024}, 2024.

\bibitem{gst}
Y.~Wang, D.~Stanton, Y.~Zhang, R.-S. Ryan, E.~Battenberg, J.~Shor, Y.~Xiao,
  Y.~Jia, F.~Ren, and R.~A. Saurous, ``Style tokens: Unsupervised style
  modeling, control and transfer in end-to-end speech synthesis,'' in {\em
  International conference on machine learning}, pp.~5180--5189, PMLR, 2018.

\bibitem{vctk}
J.~Yamagishi, C.~Veaux, and K.~MacDonald, ``Cstr vctk corpus: English
  multi-speaker corpus for cstr voice cloning toolkit (version 0.92),'' Nov
  2019.

\bibitem{ljspeech17}
K.~Ito and L.~Johnson, ``The lj speech dataset.''
  \url{https://keithito.com/LJ-Speech-Dataset/}, 2017.

\bibitem{multi_feature}
H.~Ming, L.~He, H.~Guo, and F.~Soong, ``Feature reinforcement with word
  embedding and parsing information in neural tts.,'' {\em Cornell University -
  arXiv,Cornell University - arXiv}, Jan 2019.

\bibitem{wordemb}
P.~Wang, Y.~Qian, F.~K. Soong, L.~He, and H.~Zhao, ``Word embedding for
  recurrent neural network based tts synthesis,'' in {\em 2015 IEEE
  International Conference on Acoustics, Speech and Signal Processing
  (ICASSP)}, pp.~4879--4883, 2015.

\bibitem{bert}
J.~Devlin, M.-W. Chang, K.~Lee, and K.~Toutanova, ``Bert: Pre-training of deep
  bidirectional transformers for language understanding,'' in {\em Proceedings
  of the 2019 Conference of the North}, Jan 2019.

\bibitem{g2p}
K.~Park and J.~Kim, ``g2pe.'' \url{https://github.com/Kyubyong/g2p}, 2019.

\bibitem{mos}
P.~C. Loizou, {\em Speech Quality Assessment}, pp.~623--654.
\newblock Berlin, Heidelberg: Springer Berlin Heidelberg, 2011.

\bibitem{mcd}
R.~Kubichek, ``Mel-cepstral distance measure for objective speech quality
  assessment,'' in {\em Proceedings of IEEE Pacific Rim Conference on
  Communications Computers and Signal Processing}, vol.~1, pp.~125--128 vol.1,
  1993.

\bibitem{stoi}
C.~H. Taal, R.~C. Hendriks, R.~Heusdens, and J.~Jensen, ``A short-time
  objective intelligibility measure for time-frequency weighted noisy speech,''
  in {\em 2010 IEEE International Conference on Acoustics, Speech and Signal
  Processing}, pp.~4214--4217, 2010.

\bibitem{pesq}
A.~Rix, J.~Beerends, M.~Hollier, and A.~Hekstra, ``Perceptual evaluation of
  speech quality (pesq)-a new method for speech quality assessment of telephone
  networks and codecs,'' in {\em 2001 IEEE International Conference on
  Acoustics, Speech, and Signal Processing. Proceedings (Cat. No.01CH37221)},
  vol.~2, pp.~749--752 vol.2, 2001.

\end{thebibliography}

\end{document}